\pdfoutput=1

\documentclass[11pt]{article}

\usepackage[]{ACL2023}

\usepackage{times}
\usepackage{latexsym}

\usepackage[T1]{fontenc}

\usepackage[utf8]{inputenc}

\usepackage{microtype}
\usepackage{caption}
\usepackage{subcaption}

\usepackage{inconsolata}

\usepackage{booktabs}
\usepackage{multirow}
\usepackage{graphicx}
\usepackage[normalem]{ulem}
\usepackage{rotating}
\usepackage{xparse}
\NewDocumentCommand{\avi}
{ mO{} }{\textcolor{red}{\textsuperscript{\textit{Avi}}\textsf{\textbf{\small[#1]}}}}
\newcommand{\ourcolheader}[1]{\rotatebox[origin=l]{60}{#1}}

\usepackage[]{titlesec}
\titlespacing*{\paragraph}{\parindent}{0ex}{1ex}

\title{Moving Beyond Downstream Task Accuracy for \\[0.5ex] Information Retrieval Benchmarking\thanks{{}~~Data and code will be provided as PrimeQA extensions: \url{https://github.com/primeqa/primeqa}.}}

\newcommand{\Models}{\mathcal{M}}
\newcommand{\Model}[1]{\mathcal{M}_{#1}}
\newcommand{\Metrics}{\mu}
\newcommand{\metric}{\mu}
\newcommand{\perf}{\texttt{acc}}

\newcommand{\ourauthorspace}{\hspace{4pt}}

\author{%
  Keshav Santhanam$^{1}$\Thanks{ Equal contribution.}
  \ourauthorspace
  \textbf{Jon Saad-Falcon}$^{1}$\footnotemark[2]
  \ourauthorspace
  \textbf{Martin Franz}$^{2}$
  \ourauthorspace
  \textbf{Omar Khattab}$^{1}$ 
  \ourauthorspace
  \textbf{Avirup Sil}$^{2}$\\[0.5ex]
  \ourauthorspace 
  \textbf{Radu Florian}$^{2}$
  \ourauthorspace
  \textbf{Md Arafat Sultan}$^{2}$   
  \ourauthorspace
  \textbf{Salim Roukos}$^{2}$ 
  \ourauthorspace
  \textbf{Matei Zaharia}$^{1}$ 
  \ourauthorspace 
  \textbf{Christopher Potts}$^{1}$
  \AND
  ${}^{1}$Stanford University\qquad${}^{2}$IBM Research AI}

\begin{document}
\maketitle
\begin{abstract}
Neural information retrieval (IR) systems have progressed rapidly in recent years, in large part due to the release of publicly available benchmarking tasks. Unfortunately, some dimensions of this progress are illusory: the majority of the popular IR benchmarks today focus exclusively on downstream task accuracy and thus conceal the costs incurred by systems that trade away efficiency for quality. Latency, hardware cost, and other efficiency considerations are paramount to the deployment of IR systems in user-facing settings. We propose that IR benchmarks structure their evaluation methodology to include not only metrics of accuracy, but also efficiency considerations such as a query latency and the corresponding cost budget for a reproducible hardware setting. For the popular IR benchmarks MS MARCO and XOR-TyDi, we show how the best choice of IR system varies according to how these efficiency considerations are chosen and weighed. We hope that future benchmarks will adopt these guidelines toward more holistic IR evaluation.
\end{abstract}

\section{Introduction} \label{section:introduction}

\begin{figure}[tp]
    \centering
    \includegraphics[width=0.8\columnwidth]{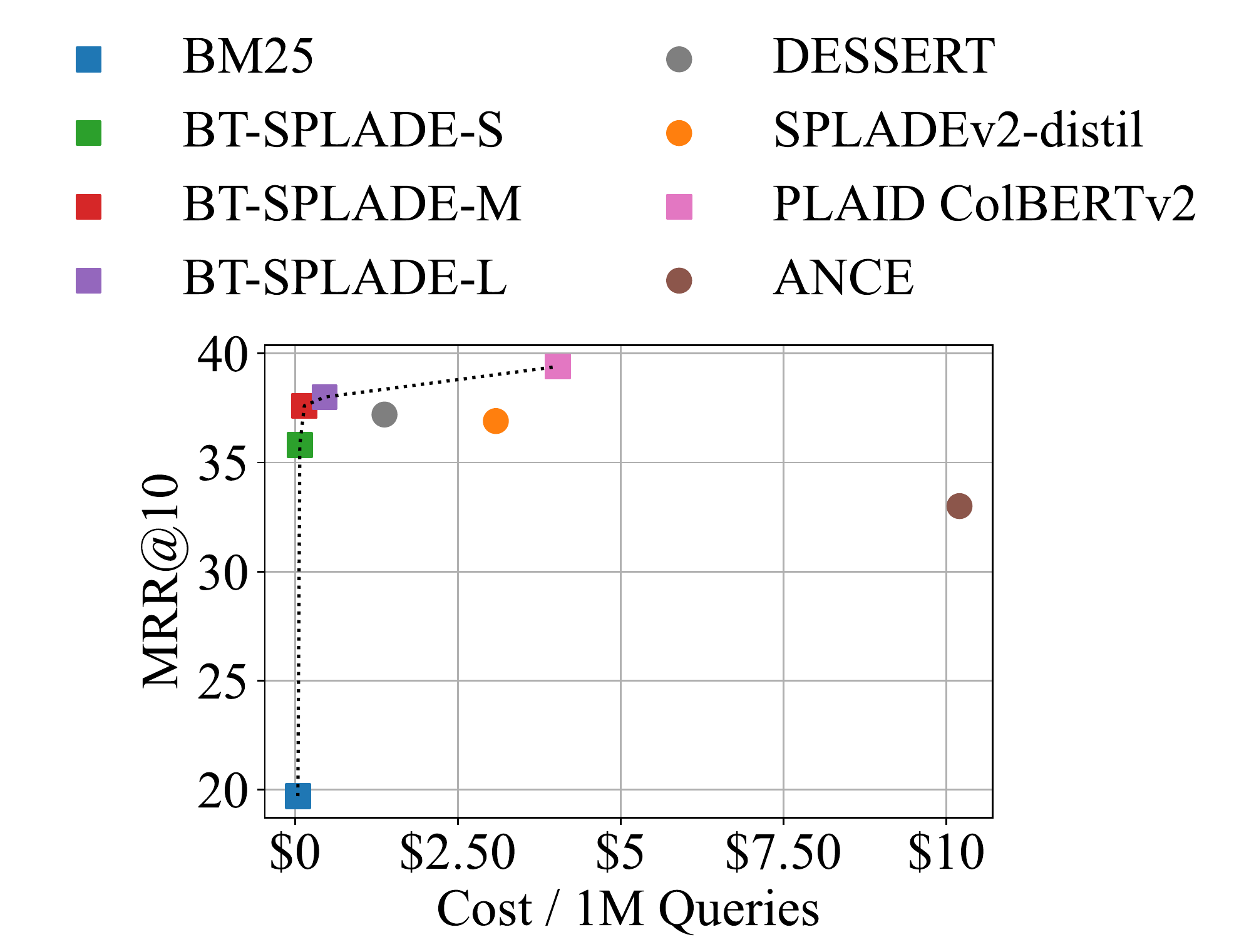}
    \caption{Selected MS MARCO Passage Ranking submissions assessed on cost and accuracy, with the Pareto frontier marked by a dotted line. The trade-offs evident here are common in real-world applications of IR technologies. These submissions do not represent ``optimal'' implementations of each respective approach, but rather reflect reported implementations and hardware configurations in the literature. Including cost and other efficiency considerations on leaderboards would lead to more thorough exploration of possible system designs and, in turn, to more meaningful progress.}
    \label{fig:pareto_frontier}
\end{figure}

Benchmark datasets have helped to drive rapid progress in neural information retrieval (IR). When the MS~MARCO \cite{nguyen2016ms} Passage Ranking leaderboard began in 2018, the best performing systems had MRR@10 scores around 0.20; the latest entries have since increased accuracy past 0.44. Similarly, the XOR~TyDi multilingual question answering (QA) dataset~\cite{asai2020xor} was released in 2021 and has seen improvements in recall scores from 0.45 to well past 0.70.

The leaderboards for these datasets are defined by a particular set of accuracy-based metrics, and progress on these metrics can easily become synonymous in people's minds with progress in general. However, IR and QA systems deployed in production environments must not only deliver high accuracy but also operate within strict resource requirements, including tight bounds on per-query latency, constraints on disk and RAM capacity, and fixed cost budgets for hardware. Within the boundaries of these constraints, the optimal solution for a downstream task may no longer be the system which simply achieves the highest task accuracy. 

\begin{table*}[tp]
\resizebox{\textwidth}{!}{
\begin{tabular}{@{} l rrr c rrr@{}}
\toprule
& \multicolumn{3}{c}{Hardware}
&& \multicolumn{3}{c}{Performance} 
\\ 
\cmidrule{2-4}  
\cmidrule{6-8} 
\multicolumn{1}{c}{}                       
& \multicolumn{1}{c}{GPU} 
& \multicolumn{1}{c}{CPU} 
& \multicolumn{1}{c}{\begin{tabular}[b]{@{}c@{}}RAM\\(GiB)\end{tabular}}   && \multicolumn{1}{c}{MRR@10} 
& \multicolumn{1}{c}{\begin{tabular}[b]{@{}c@{}}Query\\ Latency (ms)\end{tabular}} 
& \multicolumn{1}{c}{\begin{tabular}[b]{@{}c@{}}Index Size\\(GiB)\end{tabular}}  
\\
\cmidrule{1-4}  
\cmidrule{6-8} 
BM25~\cite{mackenzie2021wacky}                & 0 & 32  & 512 &                                                                                & 18.7                       & 8                                                                        &  1                                                                  \\
BM25~\cite{lassance2022efficiency}       & 0 & 64 & -  && 19.7 & 4 & 1 \\
SPLADEv2-distil~\cite{mackenzie2021wacky}    & 0 & 32 & 512 &                                                                                & 36.9                       & 220                                                                      & 4                                                                  \\
SPLADEv2-distil~\cite{lassance2022efficiency}  & 0 & 64 & - && 36.8 & 691 & 4 \\
BT-SPLADE-S~\cite{lassance2022efficiency}    & 0 & 64 & - &                                                                                  & 35.8                       & 7                                                                      &  1                                  \\
BT-SPLADE-M~\cite{lassance2022efficiency}     & 0 & 64 & - &                                                                                & 37.6                       & 13                                                                      &  2                                 \\
BT-SPLADE-L~\cite{lassance2022efficiency}    & 0 & 64 & - &                                                                               & 38.0                       & 32                                                                      &  4                                 \\   
ANCE~\cite{xiong2020approximate}             & 1 & 48 & 650 &                                                                           & 33.0                       & 12 & -                                                                         \\
RocketQAv2~\cite{ren2021rocketqav2}                           & - & - & -  &                                                                                           & 37.0                 & - &  -               \\
coCondenser~\cite{gao2021unsupervised}                          & - & - & - &                                                                                            & 38.2                 & - & -                      \\
CoT-MAE~\cite{wu2022contextual}                                   & - & - & - &                                                                        & 39.4                 & -              & -           \\
ColBERTv1~\cite{khattab2020colbert}                                              & 4 & 56 & 469 &                                                                             & 36.1                       & 54                                                                               & 154                                                                \\
PLAID ColBERTv2~\cite{Santhanam:Khattab-etal:2022}                               & 4 & 56 & 503 &                                                                               & 39.4                       & 32                                                                       &  22                                                                  \\
PLAID ColBERTv2~\cite{Santhanam:Khattab-etal:2022}                               & 4 & 56 & 503 &                                                                          & 39.4                       & 12                                                                       &  22                                                                 \\
DESSERT~\cite{engels2022dessert} & 0 & 24 & 235 &                                                                          & 37.2                       & 16                                                                       &  -  \\
 \bottomrule
\end{tabular}}
\caption{Post-hoc leaderboard of MS MARCO v1 dev performance using results reported in corresponding papers. For hardware specifications, we show the precise resources given as the running environment in the paper, even if not all resources were available to the model or the resources were over-provisioned for the particular task. Table~\ref{table:post_hoc_leaderboard_estimated_small} provides our estimates of minimum hardware requirements for a subset of these systems. Note that the first PLAID ColBERTv2 result listed was run on a server which includes 4 GPUs but no GPU was actually used for measurement, thereby resulting in a larger latency than the second listed result which does measure GPU execution.}
\label{table:post_hoc_leaderboard_reported}
\end{table*}

Figure~\ref{fig:pareto_frontier} shows how significant these trade-offs can be. The figure tracks a selection of MS~MARCO Passage Ranking submissions, with cost on the x-axis and accuracy (MRR@10) on the y-axis. At one extreme, the BM25 model costs just US\$0.04 per million queries,\footnote{Estimated by mapping the minimum necessary hardware to an AWS instance and taking the per-hour on-demand rental cost; see Table~\ref{table:post_hoc_leaderboard_estimated_small} for details.} but it is far behind the other models in accuracy. For very similar costs to BM25, one can use BT-SPLADE-S and achieve much better performance. On the other hand, the SPLADE-v2-distil model outperforms BT-SPLADE-S by about 1 point, but at a substantially higher cost. Unfortunately, these trade-offs would not be reflected on the MS~MARCO leaderboard. Similarly, the top two systems of the XOR~TyDi leaderboard as of October 2022 were separated by only 0.1 points in Recall@5000 tokens, but the gap in resource efficiency between these two approaches is entirely unclear.

In this work, we contribute to the growing literature advocating for multidimensional leaderboards that can inform different values and goals \citep{coleman2017dawnbench,mlperf,mlperf-overview,deepbench,Ma-etal:2021:Dynaboard,liu-etal-2021-explainaboard,Liang2022HolisticEO}. Our proposal is that researchers should report orthogonal dimensions of performance such as query latency and overall cost, in addition to  accuracy-based metrics. Our argument has two main parts. 

In part~1 (\S\ref{sec:post-hoc}), we create a post-hoc MS~MARCO leaderboard from published papers (Table~\ref{table:post_hoc_leaderboard_reported}). This reveals that systems with similar accuracy often differ substantially along other dimensions, and also that techniques for improving latency and reducing memory and hardware costs are currently being explored only very sporadically. However, a few of the contributions \citep{Santhanam:Khattab-etal:2022,lassance2022efficiency,engels2022dessert,li2022citadel} exemplify the kind of thorough investigation of accuracy and efficiency that we are advocating for, and we believe that improved multidimensional leaderboards could spur further innovation in these areas. 

In part~2 (\S\ref{sec:experimental}), we systematically explore four prominent systems: BM25, Dense Passage Retriever (DPR; \citealt{karpukhin-etal-2020-dense}), \mbox{BT-SPLADE-L}~\citep{formal2021splade,lassance2022efficiency}, and PLAID ColBERTv2 \citep{khattab2020colbert,Santhanam:Khattab-etal:2022,santhanam-etal-2022-colbertv2}. These experiments begin to provide a fuller picture of the overall performance of these systems. 

We close by discussing practical considerations relating to the multidimensional leaderboards that the field requires. Here, we argue that the \emph{Dynascore} metric developed by \citet{Ma-etal:2021:Dynaboard} is a promising basis for leaderboards that aim to (1) measure systems along multiple dimensions and (2) provide a single full ranking of systems. Dynascores allow the leaderboard creator to weight different assessment dimensions (e.g., to make cost more important than latency). These weightings transparently reflect a particular set of values, and we show that they give rise to leaderboards that are likely to incentivize different research questions and system development choices than current leaderboards do.
\section{A Post-hoc Leaderboard}\label{sec:post-hoc}

While existing IR benchmarks facilitate progress on accuracy metrics, the lack of a unified methodology for measuring latency, memory usage, and hardware cost makes it challenging to understand the trade-offs between systems. To illustrate this challenge, we constructed a post-hoc leaderboard for the MS~MARCO Passage Ranking benchmark (Table~\ref{table:post_hoc_leaderboard_reported}). We include the MRR@10 values reported in prior work and, when available, copy the average per-query latency, index size, and hardware configurations reported in the respective papers.\footnote{We plan to expand our analysis to include the recently released CITADEL model \cite{li2022citadel}, first uploaded to arXiv on 11/18/22)} We highlight the following key takeaways.

\begin{table}[tp]
\footnotesize
\setlength{\tabcolsep}{4pt}
\resizebox{\linewidth}{!}{
\begin{tabular}[b]{ @{} l@{ } *{3}{r} l r @{}}
\toprule                     
& GPU
& CPU
& RAM
& Instance
& Cost
\\ \midrule 
BM25             & 0 & 1 & 4 & m6g.med & \$0.04 \\
SPLADEv2-distil  & 0 & 1 & 8 & r6g.med & \$3.08 \\
BT-SPLADE-S      & 0 & 1 & 8 & m6g.med & \$0.07 \\
BT-SPLADE-M      & 0 & 1 & 8 & m6g.med &  \$0.14 \\
BT-SPLADE-L      & 0 & 1 & 8 & r6g.med & \$0.45 \\   
ANCE             & 1 & 8 & 64 &  p3.2xl & \$10.20 \\
ColBERTv1 & 1 & 16 & 256 &  p3.8xl  & \$183.60 \\
PLAID ColBERTv2 & 0 & 8 & 32 &  r6a.2xl &  \$4.03 \\
PLAID ColBERTv2 & 1 & 8 & 64 &  p3.2xl  &  \$10.20 \\
DESSERT & 0 & 8 & 32 & m6g.2xl  & \$1.37 \\
 \bottomrule
\end{tabular}
}
\caption{Estimated minimum viable AWS instance type necessary to run each model. RAM is in GiB; Cost is per 1M queries.}
\label{table:post_hoc_leaderboard_estimated_small}
\end{table}

\subsection{Hardware Provisioning}

The hardware configurations in Table~\ref{table:post_hoc_leaderboard_reported} are the specific compute environments listed in the corresponding papers rather than the minimum viable hardware necessary to achieve the reported latency. In Table~\ref{table:post_hoc_leaderboard_estimated_small}, we have sought to specify the minimal configuration that would be needed to run each system. (This may result in an overly optimistic assessment of latency; see \S\ref{sec:experimental}). The hardware differences between Table~\ref{table:post_hoc_leaderboard_reported} and Table~\ref{table:post_hoc_leaderboard_estimated_small} reveal that researchers are often using vastly over-provisioned hardware for their experiments. Our proposed leaderboards would create a pressure to be more deliberative about the costs of hardware used when reporting efficiency metrics.

\subsection{Variation in Methodology} 

Table~\ref{table:post_hoc_leaderboard_reported} shows that both the quality metrics and the hardware used for evaluation across different models vary significantly. Many papers exclusively report accuracy, which precludes any quantitative understanding of efficiency implications~\cite{ren2021rocketqav2, gao2021unsupervised, wu2022contextual}. For papers that do report efficiency-oriented metrics, the evaluation environment and methodology are often different; for example, the results from \citealt{mackenzie2021wacky} and \citealt{lassance2022efficiency} are measured on a single CPU thread whereas \citealt{khattab2020colbert} and \citealt{Santhanam:Khattab-etal:2022} leverage multiple CPU threads for intra-query parallelism, and even a GPU for certain settings. We also observe performance variability even for the same model, with \citealt{mackenzie2021wacky} (220 ms) and \citealt{lassance2022efficiency} (691 ms) reporting SPLADEv2 latency numbers which are 3$\times$ apart. Similarly, the BM25 latencies reported by these papers differ by a factor of 2$\times$. 

\subsection{Multidimensional Evaluation Criteria} 

The optimal model choice for MS~MARCO is heavily dependent on how we weight the different evaluation metrics. Based purely on accuracy, CoT-MAE and PLAID ColBERTv2 are the top-performers in Table~\ref{table:post_hoc_leaderboard_reported}, with an MRR@10 score of 39.4 for both. However, we do not have all the information we need to compare them along other dimensions. On the other hand, BM25 is the fastest model, with a per-query latency of only 4~ms as measured by \citet{lassance2022efficiency}, and its space footprint is also small. The trade-off is that it has the lowest accuracy in the cohort. Compared to BM25, one of the highly optimized BT-SPLADE models may be a better choice. Figure~\ref{fig:pareto_frontier} begins to suggest how we might reason about these often opposing pressures.

\section{Experiments with Representative Retrievers} \label{sec:experimental}

\begin{table*}[ht]
\begin{minipage}[c]{0.48\textwidth}
\begin{subtable}[c]{1\textwidth}
\resizebox{\linewidth}{!}{
\begin{tabular}[t]{@{}l rrrr c  rr@{}}
\toprule
& \multicolumn{4}{c}{Hardware}
&
& \multicolumn{2}{c}{Performance}
\\ 
\cmidrule(l){2-5}    
\cmidrule(l){7-8} 
& \ourcolheader{GPU}
& \ourcolheader{CPU}
& \ourcolheader{RAM}
& \ourcolheader{Instance}
&
& \ourcolheader{Latency}
& \ourcolheader{Cost}
\\ 
\cmidrule{1-5}\cmidrule{7-8} 
BM25                                        
& \multirow{1}{*}{0}      
& \multirow{1}{*}{1}      
& \multirow{1}{*}{4}            
& \multirow{1}{*}{m6gd.med} &                  
& 11     
& \$0.14                                                          
\\

\cmidrule{1-5}\cmidrule{7-8} 
BM25                                        
& \multirow{6}{*}{0}      
& \multirow{6}{*}{1}      
& \multirow{6}{*}{32}           
& \multirow{6}{*}{x2gd.lrg}  &                    
& 10                     
& \$0.48                            
\\
DPR                                         
&                         
&                        
&                                                              
&&                                        
& 146                    
& \$6.78                            
\\
ColBERTv2-S                             
&                         
&                         
&                                                                  
&&                                        
& 206                     
& \$9.58                            
\\
ColBERTv2-M                             
&                         
&                         
&  

&&                                        
& 321                        
& \$14.90                                   
\\
ColBERTv2-L                             
&                         
&                         
&  

&&                                        
& 459                        
& \$21.30                                   
\\
BT-SPLADE-L                             
&                         
&                         
&  

&&                                        
& 46                        
& \$2.15                                   
\\
\cmidrule{1-5}\cmidrule{7-8} 
BM25                                        
& \multirow{1}{*}{0}      
& \multirow{1}{*}{16}     
& \multirow{1}{*}{4}            
& \multirow{1}{*}{c7g.4xl}  &                 
& 9                     
& \$1.48                            
\\

\cmidrule{1-5}\cmidrule{7-8} 
BM25                                        
& \multirow{6}{*}{0}      
& \multirow{6}{*}{16}     
& \multirow{6}{*}{32}           
& \multirow{6}{*}{c7g.4xl}  &            
& 9                      
& \$1.43                            
\\
DPR                                         
&                         
&                         
&                                                                 
&&                                        
& 19                   
& \$2.97                            
\\
ColBERTv2                            
&                         
&                        
&                                                               
&&                                        
& 51                     
& \$8.19                            
\\
ColBERTv2-M                             
&                         
&                         
&  

&&                                        
& 63                        
& \$10.09                                   
\\
ColBERTv2-L                             
&                         
&                         
&  

&&                                        
& 86                        
& \$13.88                                   
\\
BT-SPLADE-L                             
&                         
&                         
&                                                                 
&&                                       
& 33                         
& \$5.38                                   
\\
\cmidrule{1-5}\cmidrule{7-8} 
BM25                                        
& \multirow{1}{*}{1}      
& \multirow{1}{*}{1}      
& \multirow{1}{*}{4}            
& \multirow{1}{*}{p3.2xl}  &                   
& 11                    
& \$9.09                           
\\

\cmidrule{1-5}\cmidrule{7-8} 
BM25                                        
& \multirow{6}{*}{1}      
& \multirow{6}{*}{1}      
& \multirow{6}{*}{32}           
& \multirow{6}{*}{p3.2xl}  &                     
& 10                     
& \$8.46                            
\\
DPR                                         
&                         
&                         
&                                                                  
&&                                       
& 19                    
& \$15.73                            
\\
ColBERTv2-S                            
&                         
&                         
&                                                                  
&&                                       
& 36                     
& \$30.46                            
\\
ColBERTv2-M                             
&                         
&                         
&  

&&                                        
& 52                        
& \$44.54                                   
\\
ColBERTv2-L                             
&                         
&                         
&  

&&                                        
& 99                        
& \$83.97                                   
\\
BT-SPLADE-L                            
&                         
&                         
&                                                                 
&&                                        
& 42                        
& \$35.86                                   
\\
\cmidrule{1-5}\cmidrule{7-8} 
BM25                                        
& \multirow{1}{*}{1}      
& \multirow{1}{*}{16}     
& \multirow{1}{*}{4}            
& \multirow{1}{*}{p3.8xl} &                     
& 9                     
& \$30.51                            
\\

\cmidrule{1-5}\cmidrule{7-8} 
BM25                                       
& \multirow{6}{*}{1}      
& \multirow{6}{*}{16}     
& \multirow{6}{*}{32}           
& \multirow{6}{*}{p3.8xl}  &                  
& 9         
& \$29.94                            
\\
DPR                                        
&                         
&                        
&                                                                 
&&                                        
& 18                  
& \$61.06                            
\\
ColBERTv2-S                            
&                         
&                         
&                                                                  
&&                                        
& 27                     
& \$90.41                            
\\
ColBERTv2-M                             
&                         
&                         
&  

&&                                        
& 36                        
& \$123.35                                   
\\
ColBERTv2-L                             
&                         
&                         
&  

&&                                        
& 55                        
& \$187.24                                   
\\
BT-SPLADE-L                             
&                         
&                         
&                                                                 
&&                                        
& 33                        
& \$112.87                                   
\\
\bottomrule
\end{tabular}}
\caption{MS MARCO efficiency results.}
\label{table:msmarco_perf}
\end{subtable}
\end{minipage}
\hfill
\begin{minipage}[c]{0.48\textwidth}
\begin{subtable}[b]{1\linewidth}
\resizebox{\columnwidth}{!}{
\begin{tabular}[t]{@{}l rrrr c rr@{}}
\toprule
& \multicolumn{4}{c}{Hardware}
&
& \multicolumn{2}{c}{Performance}
\\ 
\cmidrule(l){2-5}\cmidrule(l){7-8} 
& \ourcolheader{GPU}
& \ourcolheader{CPU}
& \ourcolheader{RAM}
& \ourcolheader{Instance}
&
& \ourcolheader{Latency}
& \ourcolheader{Cost}
\\ 
\cmidrule{1-5}\cmidrule{7-8} 
BM25                                        
& \multirow{6}{*}{0}      
& \multirow{6}{*}{1}      
& \multirow{6}{*}{64}                                                 & \multirow{6}{*}{x2gd.xlrg}    &                 
& 37                    
& \$3.45                            
\\
DPR                                         
&                         
&                         
& 
&&
& 208               
& \$19.29                     
\\
ColBERTv2-S                             
&                         
&                         
&                                                                     
&&
& 343
& \$31.84   
\\
ColBERTv2-M                             
&                         
&                         
&                                                                     
&&
& 771 
& \$71.56
\\
ColBERTv2-L                             
&                         
&                         
&                                                                     
&&
& 1107 
& \$102.74   
\\
BT-SPLADE-L                            
&                         
&                         
&                                                                     
&&
& 70 
& \$6.49                            
\\
\cmidrule{1-5}\cmidrule{7-8} 
BM25                                        
& \multirow{6}{*}{0}      
& \multirow{6}{*}{16}     
& \multirow{6}{*}{64} 
& \multirow{6}{*}{m6g.4xlrg} &     
& 36                  
& \$6.11                            
\\
DPR                                         
&                                                  
&  
&                                   
& &                                       
& 84                 
& \$14.38                            
\\
ColBERTv2-S                      
&                                                  
&                                                                     &                                   
&    &                                    
& 83        
& \$14.17                                 
\\
ColBERTv2-M                             
&                         
&                         
&                                                                     
&&
& 110 
& \$18.83
\\
ColBERTv2-L                             
&                         
&                         
&                                                                     
&&
& 165 
& \$28.26   
\\
BT-SPLADE-L                            
&                         
&                         
&                                                                     
&&
& 43 
& \$7.41                            
\\
\cmidrule{1-5}\cmidrule{7-8} 
BM25                                        
& \multirow{6}{*}{1}      
& \multirow{6}{*}{1}      
& \multirow{6}{*}{64}                                                  & \multirow{6}{*}{p3.8xl}  &                     
& 36            
& \$123.69                            
\\
DPR                                         
&                                                  
&  
&                                   
&&                                    
& 26                
& \$89.91                           
\\
ColBERTv2-S                             
&                                                  
&  
&                                   
&&                                    
& 57                 
& \$194.81                            
\\
ColBERTv2-M                             
&                         
&                         
&                                                                     
&&
& 74 
& \$251.76
\\
ColBERTv2-L                             
&                         
&                         
&                                                                     
&&
& 121 
& \$411.62   
\\
BT-SPLADE-L                            
&                         
&                         
&                                                                     
&&
& 63 
& \$213.17                           
\\
\cmidrule{1-5}\cmidrule{7-8} 
BM25                                        
& \multirow{6}{*}{1}      
& \multirow{6}{*}{16}     
& \multirow{6}{*}{64}                                                 & \multirow{6}{*}{p3.8xl} &                      
& 35                   
& \$118.12                            
\\
DPR                                         
&                         
&                         
&   
& &
& 28 
& \$95.23                           
\\
ColBERTv2-S                             
&                                                  
&                                                                     &                                   
&&                                        
& 46                    
& \$155.10                           
\\
ColBERTv2-M                             
&                         
&                         
&                                                                     
&&
& 65
& \$219.84
\\
ColBERTv2-L                             
&                         
&                         
&                                                                     
&&
& 106 
& \$359.65   
\\
BT-SPLADE-L                            
&                         
&                         
&                                                                     
&&
& 43 
& \$147.53                            
\\
\bottomrule 
\end{tabular}
}
\caption{XOR-TyDi efficiency results.}
\label{table:xor_tydi_perf}
\end{subtable}

\begin{subtable}{1\linewidth}
\resizebox{\columnwidth}{!}{
\begin{tabular}{l r r r r}
\toprule
& \multicolumn{2}{c}{MS MARCO} 
& \multicolumn{2}{c}{XOR-TyDi} \\
& MRR@10 
& Success@10 
& MRR@10 
& Success@10 \\
\midrule     
BM25 &  18.7 & 38.6 & 26.3 & 44.5\\
DPR  &  31.7 & 52.1 & 16.9 & 32.4 \\
ColBERTv2-S & 39.4 & 68.8  & 41.8 &  57.5 \\
ColBERTv2-M & 39.7 & 69.6  & 45.4 & 63.0 \\
ColBERTv2-L & 39.7 & 69.7  & 47.4 & 66.0 \\
BT-SPLADE-L & 38.0 & 66.3 & 43.5 & 65.4 \\
\bottomrule
\end{tabular}
}
\caption{Accuracy.}
\label{tab:accuracy}
\end{subtable}
\end{minipage}

\vspace{-8pt}

\caption{Experimental results. Latency is average per-query latency (ms), and Cost is per 1M queries.}
\label{tab:results}
\end{table*}

As Table~\ref{table:post_hoc_leaderboard_reported} makes clear, the existing literature does not include systematic, multidimensional comparisons of models. In this section, we report on experiments that allow us to make these comparisons. We focus on four models:
\paragraph{BM25 \citep{robertson1995okapi}} A sparse, term-based IR model. BM25 remains a strong baseline in many IR contexts and is notable for its low latency and low costs. We assess a basic implementation. More sophisticated versions may achieve better accuracy \citep{berger1999information,boytsov2020traditional}, though often with trade-offs along other dimensions \citep{lin2016toward}. For evidence that simple BM25 models often perform best in their class, see \citealt{nandan2021beir}.

\paragraph{DPR \citep{karpukhin-etal-2020-dense}} A dense single-vector neural IR model. DPR separately encodes queries and documents into vectors and scores them using fast dot-product-based comparisons.

\paragraph{BT-SPLADE-L \citep{lassance2022efficiency}} SPLADE  \citep{formal2021splade} is a sparse neural model. The BT-SPLADE variants are highly optimized versions of this model designed to achieve low latency and reduce the overall computational demands of the original model. To the best of our knowledge, only the Large configuration, BT-SPLADE-L, is publicly available.

\paragraph{PLAID ColBERTv2 \citep{Santhanam:Khattab-etal:2022}} The ColBERT retrieval model~\cite{khattab2020colbert} encodes queries and documents into sequences of output states, one per input token, and scoring is done based on the maximum similarity values obtained for each query token. ColBERTv2~\cite{santhanam-etal-2022-colbertv2} improves supervision and reduces the space footprint of the index, and the PLAID engine focuses on achieving low latency. The parameter $k$ to the model dictates the initial candidate passages that are scored by the model. Larger $k$ thus leads to higher latency but generally more accurate search. In our initial experiments, we noticed that higher $k$ led to better out-of-domain performance, and thus we evaluated the recommended settings from \citet{Santhanam:Khattab-etal:2022}, namely, $k \in \{10, 100, 1000\}$. To distinguish these configurations from the number of passages evaluated by the MRR or Success metric (also referred to as $k$), we refer to these configurations as the `-S', `-M', and `-L' variants of ColBERTv2, respectively.

We chose these models as representatives of key IR model archetypes: lexical models (BM25), dense single-vector models (DPR), sparse neural models (SPLADE), and late-interaction models (ColBERT). The three ColBERT variants provide a glimpse of how model configuration choices can interact with our metrics.

We use two retrieval datasets: MS~MARCO \cite{nguyen2016ms} and XOR-TyDi \cite{asai2020xor}. All neural models in our analysis are trained on MS~MARCO data. We evaluate on XOR-TyDi without further fine-tuning to test out-of-domain evaluation (see Appendix~\ref{app:experiment_details} for more details).

Our goal is to understand how the relative performance of these models changes depending on the available resources and evaluation criteria. Our approach differs from the post-hoc leaderboard detailed in \S\ref{sec:post-hoc} in two key ways: (1) we fix the underlying hardware platform across all models, and (2) we evaluate each model across a broad range of hardware configurations (AWS instance types), ensuring that we capture an extensive space of compute environments. Furthermore, in addition to quality, we also report the average per-query latency and the corresponding cost of running 1 million queries given the latency and the choice of instance type. This approach therefore enables a more principled and holistic comparison between the models.

We use the open-source PrimeQA framework,\footnote{\url{https://github.com/primeqa/primeqa}} which provides a uniform interface to implementations of BM25, DPR, and PLAID ColBERTv2. For SPLADE, we use the open-source implementation maintained by the paper authors.\footnote{\url{https://github.com/naver/splade}} For each model we retrieve the top 10 most relevant passages. We report the average latency of running a fixed sample of 1000 queries from each dataset as measured across 5 trials. See Appendix~\ref{app:experiment_details} for more details about the evaluation environments and model configurations. 

Table~\ref{tab:results} summarizes our experiments. Tables~\ref{table:msmarco_perf} and \ref{table:xor_tydi_perf} report efficiency numbers, with costs estimated according to the same hardware pricing used for Table~\ref{table:post_hoc_leaderboard_estimated_small}.  Table~\ref{tab:accuracy} gives accuracy results (MRR@10 and Success@10).

Overall, BM25 is the least expensive model when selecting the minimum viable instance type: only BM25 is able to run with 4 GB memory. However, its accuracy scores are low enough to essentially remove it from contention.

On both datasets, we find that BT-SPLADE-L and the PLAID ColBERTv2 variants are the most accurate models by considerable margins. On MS MARCO, all the ColBERTv2 variants outperform BT-SPLADE-L in MRR@10 and Success@10 respectively, while BT-SPLADE-L offers faster and cheaper scenarios than ColBERTv2 for applications that permit a moderate loss in quality. 

In the out-of-domain XOR-TyDi evaluation, BT-SPLADE-L outperforms the ColBERTv2-S variant, which sets $k=10$ (the least computationally-intensive configuration). We hypothesize this loss in quality is an artifact of the approximations employed by the default configuration. Hence, we also test the more computationally-intensive configurations mentioned above: ColBERTv2-M ($k = 100$) and ColBERTv2-L ($k = 1000$). These tests reveals that ColBERTv2-L solidly outperforms BT-SPLADE-L in MRR@10 and Success@10, while allowing BT-SPLADE-L to expand its edge in latency and cost. 

Interestingly, despite per-instance costs being higher for certain instances, selecting the more expensive instance can actually reduce cost depending on the model. For example, the \texttt{c7g.4xlarge} instance is 3.5$\times$ more expensive than \texttt{x2gd.large}, but ColBERTv2-S runs 4$\times$ faster with 16 CPU threads and therefore is cheaper to execute on the \texttt{c7g.4xlarge}. These findings further reveal the rich space of trade-offs when it comes to model configurations, efficiency, and accuracy.

\section{Discussion and Recommendations} \label{sec:discussion}

In this section, we highlight several considerations for future IR leaderboards and offer recommendations for key design decisions.

\subsection{Evaluation Platform} \label{sec:discussion_evaluation_platform} 

A critical design goal for IR leaderboards should be to encourage transparent,  reproducible submissions. However, as we see in Table~\ref{table:post_hoc_leaderboard_reported}, many existing submissions are performed using custom---and likely private---hardware configurations and are therefore difficult to replicate.

Instead, we strongly recommend all submissions be tied to a particular public cloud instance type.\footnote{In principle, any public cloud provider (e.g., AWS EC2, Google Cloud, or Azure) is acceptable as long as they offer a transparent way to estimate costs.} In particular, leaderboards should require that the specific evaluation environment associated with each submission (at inference time) can be easily reproduced. This encourages submissions to find realistic and transparent ways to use public cloud resources that minimize the cost of their submissions in practice, subject to their own goals for latency and quality. We note that our inclusion of ``cost'' subsumes many individual tradeoffs that systems may consider, like the amount of RAM (or, in principle, storage) required by the index and model, or the number of CPUs, GPUs, or TPUs.

In principle, leaderboards could report the constituent resources instead of reporting a specific reproducible hardware platform. For example, a leaderboard could simply report the number of CPU threads and GPUs per submission. This offers the benefit of decoupling submissions from the offerings available on public cloud providers. However, this approach fails to account for the ever-growing space of hardware resources or their variable (and changing) pricing. For instance, it is likely unrealistic to expect leaderboard builders to quantify the difference in cost between a V100 and a more recent A100 GPU---or newer generations, like H100, let alone FPGAs or other heterogeneous choices. We argue that allowing submissions to select their own public cloud instance (including its capabilities and pricing) reflects a realistic, market-driven, up-to-date strategy for estimating dollar costs. In practice, the leaderboard creators need to set a policy for dealing with changing prices over time. They may, for instance, opt to use the latest pricing at all times. This may lead to shifts in the leaderboard rankings over time, reflecting the changing tradeoffs between cost and the other dimensions evaluated.

\subsection{Scoring} \label{sec:discussion_scoring}

\begin{table*}[ht]
    \begin{subtable}{0.48\linewidth}
        \resizebox{\columnwidth}{!}{
            \begin{tabular}{lllr}
\toprule
{} &       System &                     Hardware &  Dynascore \\
\midrule
1  &  ColBERTv2-M &         16 CPU, 32 GB memory &     19.127 \\
2  &  ColBERTv2-S &         16 CPU, 32 GB memory &     19.118 \\
3  &  ColBERTv2-L &         16 CPU, 32 GB memory &     18.857 \\
4  &  ColBERTv2-S &   1 GPU, 1 CPU, 32 GB memory &     18.698 \\
5  &  BT-SPLADE-L &         16 CPU, 32 GB memory &     18.637 \\
6  &  BT-SPLADE-L &          1 CPU, 32 GB memory &     18.616 \\
7  &  ColBERTv2-M &   1 GPU, 1 CPU, 32 GB memory &     18.385 \\
8  &  ColBERTv2-S &          1 CPU, 32 GB memory &     17.912 \\
9  &  BT-SPLADE-L &   1 GPU, 1 CPU, 32 GB memory &     17.839 \\
10 &  ColBERTv2-S &  1 GPU, 16 CPU, 32 GB memory &     17.331 \\
11 &  ColBERTv2-L &   1 GPU, 1 CPU, 32 GB memory &     17.080 \\
12 &  ColBERTv2-M &          1 CPU, 32 GB memory &     17.060 \\
13 &  ColBERTv2-M &  1 GPU, 16 CPU, 32 GB memory &     16.619 \\
14 &  BT-SPLADE-L &  1 GPU, 16 CPU, 32 GB memory &     16.062 \\
15 &  ColBERTv2-L &          1 CPU, 32 GB memory &     15.858 \\
16 &          DPR &         16 CPU, 32 GB memory &     15.635 \\
17 &          DPR &   1 GPU, 1 CPU, 32 GB memory &     15.330 \\
18 &  ColBERTv2-L &  1 GPU, 16 CPU, 32 GB memory &     14.940 \\
19 &          DPR &          1 CPU, 32 GB memory &     14.583 \\
20 &          DPR &  1 GPU, 16 CPU, 32 GB memory &     14.252 \\
21 &         BM25 &           1 CPU, 4 GB memory &      9.263 \\
22 &         BM25 &          1 CPU, 32 GB memory &      9.263 \\
23 &         BM25 &         16 CPU, 32 GB memory &      9.248 \\
24 &         BM25 &          16 CPU, 4 GB memory &      9.246 \\
25 &         BM25 &   1 GPU, 1 CPU, 32 GB memory &      9.072 \\
26 &         BM25 &    1 GPU, 1 CPU, 4 GB memory &      9.049 \\
27 &         BM25 &  1 GPU, 16 CPU, 32 GB memory &      8.565 \\
28 &         BM25 &   1 GPU, 16 CPU, 4 GB memory &      8.551 \\
\bottomrule
\end{tabular}

    }
    \caption{MS MARCO.}
    \end{subtable}
    \hfill
    \begin{subtable}{0.48\linewidth}
        \resizebox{\columnwidth}{!}{
            \begin{tabular}{lllr}
\toprule
{} &       System &                     Hardware &  Dynascore \\
\midrule
1  &  ColBERTv2-L &         16 CPU, 64 GB memory &     21.241 \\
2  &  BT-SPLADE-L &         16 CPU, 64 GB memory &     21.119 \\
3  &  ColBERTv2-M &         16 CPU, 64 GB memory &     21.063 \\
4  &  BT-SPLADE-L &          1 CPU, 64 GB memory &     20.753 \\
5  &  ColBERTv2-M &  1 GPU, 16 CPU, 64 GB memory &     20.255 \\
6  &  BT-SPLADE-L &  1 GPU, 16 CPU, 64 GB memory &     20.123 \\
7  &  ColBERTv2-M &   1 GPU, 1 CPU, 64 GB memory &     19.904 \\
8  &  ColBERTv2-L &  1 GPU, 16 CPU, 64 GB memory &     19.700 \\
9  &  ColBERTv2-S &         16 CPU, 64 GB memory &     19.649 \\
10 &  BT-SPLADE-L &   1 GPU, 1 CPU, 64 GB memory &     19.380 \\
11 &  ColBERTv2-S &  1 GPU, 16 CPU, 64 GB memory &     19.157 \\
12 &  ColBERTv2-L &   1 GPU, 1 CPU, 64 GB memory &     19.123 \\
13 &  ColBERTv2-S &   1 GPU, 1 CPU, 64 GB memory &     18.723 \\
14 &  ColBERTv2-S &          1 CPU, 64 GB memory &     15.934 \\
15 &         BM25 &          1 CPU, 64 GB memory &     12.635 \\
16 &         BM25 &         16 CPU, 64 GB memory &     12.630 \\
17 &         BM25 &  1 GPU, 16 CPU, 64 GB memory &     11.847 \\
18 &         BM25 &   1 GPU, 1 CPU, 64 GB memory &     11.794 \\
19 &  ColBERTv2-M &          1 CPU, 64 GB memory &     11.563 \\
20 &  ColBERTv2-L &          1 CPU, 64 GB memory &      7.708 \\
21 &          DPR &   1 GPU, 1 CPU, 64 GB memory &      7.452 \\
22 &          DPR &  1 GPU, 16 CPU, 64 GB memory &      7.386 \\
23 &          DPR &         16 CPU, 64 GB memory &      7.188 \\
24 &          DPR &          1 CPU, 64 GB memory &      5.442 \\
\bottomrule
\end{tabular}

        }
    \caption{XOR-TyDi.}
    \end{subtable}
\caption{Dynascores for the default weighting scheme \{Accuracy:~0.5, Cost:~0.25, Latency:~0.25\}.}
\label{tab:dynascore-leaderboard}
\end{table*}

Efficiency-aware IR leaderboards have several options for scoring and ranking submissions. We enumerate three such strategies here: 
\begin{enumerate}\setlength{\itemsep}{0pt}
    \item Fix a latency or cost threshold (for example) and rank eligible systems by accuracy. Many different thresholds could be chosen to facilitate competition in different resource regimes (e.g., mobile phones vs.~data centers). 
    \item Fix an accuracy threshold and rank eligible systems by latency or cost (or other aspects). The accuracy threshold could be set to the state-of-the-art result from prior years.
    \item Weight the different assessment dimensions and distill them into a single score, possibly after filtering systems based on thresholds on accuracy, latency, and/or cost.
\end{enumerate}

Of these approaches, the third is the most flexible and is the only one that can provide a complete ranking of systems. The \emph{Dynascores} of \citet{Ma-etal:2021:Dynaboard} seem particularly well-suited to IR leaderboards, since they allow the leaderboard creator to assign weights to each of the dimensions included in the assessment, reflecting the relative importance assigned to each. The Dynascore itself is
a utility-theoretic aggregation of all the measurements and yields a ranking of the systems under consideration.  

Following \citeauthor{Ma-etal:2021:Dynaboard}, we define Dynascores as follows. For a set of models $\Models = \{\Model{i}, \ldots, \Model{N}\}$ and assessment metrics $\Metrics = \{\metric_{1}, \ldots, \metric_{k}\}$, the Dynascore for a model $\Model{i} \in \Models$ is defined as
\begin{equation}
\sum_{j=1}^{k}
\mathbf{w}_{\metric_{j}}
\frac{
    \metric_{j}(\Model{i})
}{
    \texttt{AMRS}(\Models, \perf, \metric_{j})
}
\end{equation}
where $\mathbf{w}_{\metric_{j}}$ is the weight assigned to $\metric_{j}$ (we ensure that the sum of all the weights is equal to $1$), and $\perf$ is an appropriate notion of accuracy (e.g., MRR@10). The \texttt{AMRS} is defined as
\begin{equation}
    \frac{1}{N}
    \sum_{i}^{N}
    \left|
    \frac{
        \metric(\Model{i}) - \metric(\Model{i+1})
    }{
        \perf(\Model{i}) - \perf(\Model{i+1})
    }
    \right|
\end{equation}
for models $\Model{i}, \ldots, \Model{N}$ organized from worst to best performing according to $\perf$.
In our experiments, we use the negative of Cost and Latency, so that all the metrics are oriented in such a way that larger values are better. If a model cannot be run for a given hardware configuration, it is excluded.

\begin{figure*}[tp]
\begin{subfigure}{0.46\textwidth}
\centering
\includegraphics[width=1\textwidth]{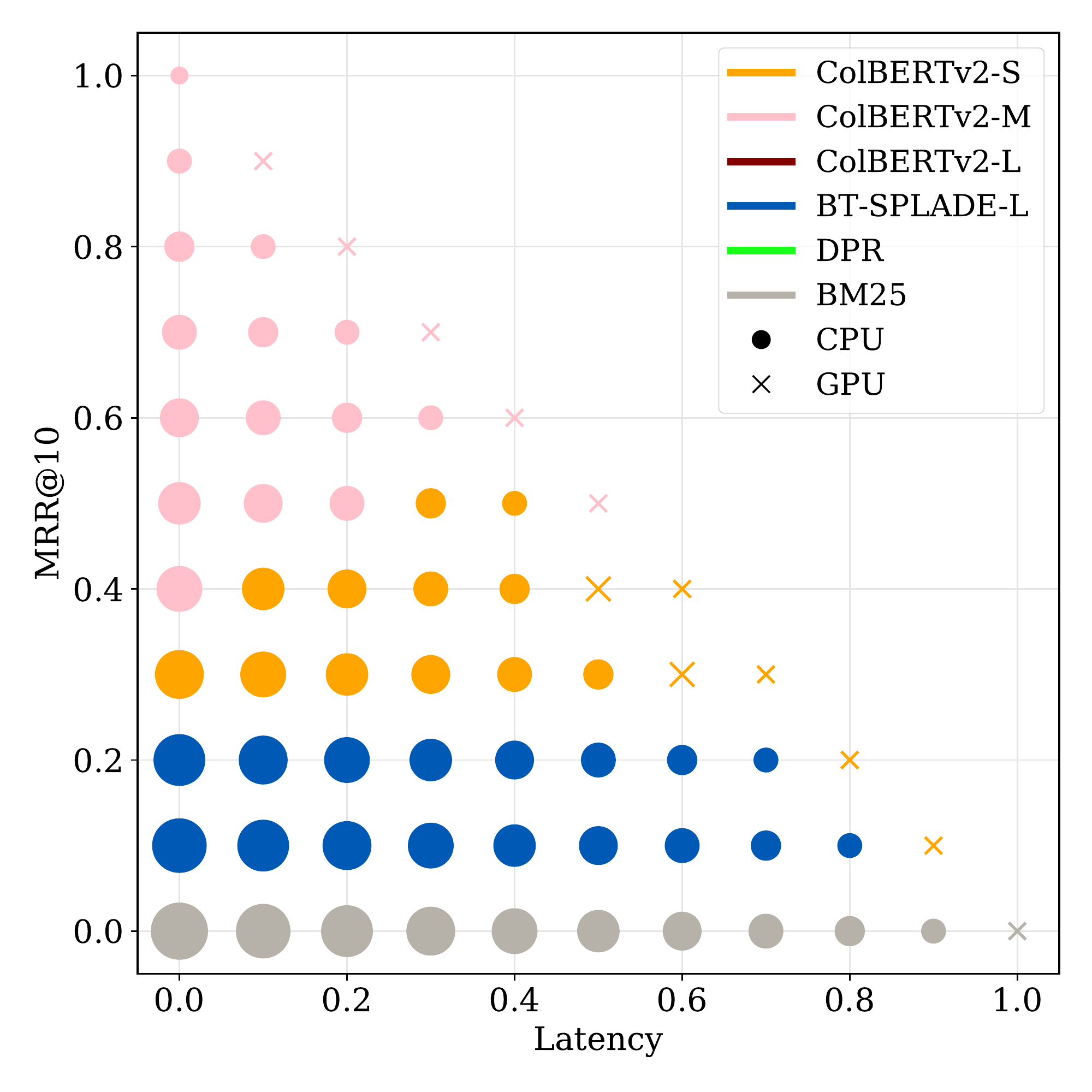}
\caption{MS~MARCO.}
\label{fig:dynascore-full}
\end{subfigure}
\hfill
\begin{subfigure}{0.46\textwidth}
\centering
\includegraphics[width=1\textwidth]{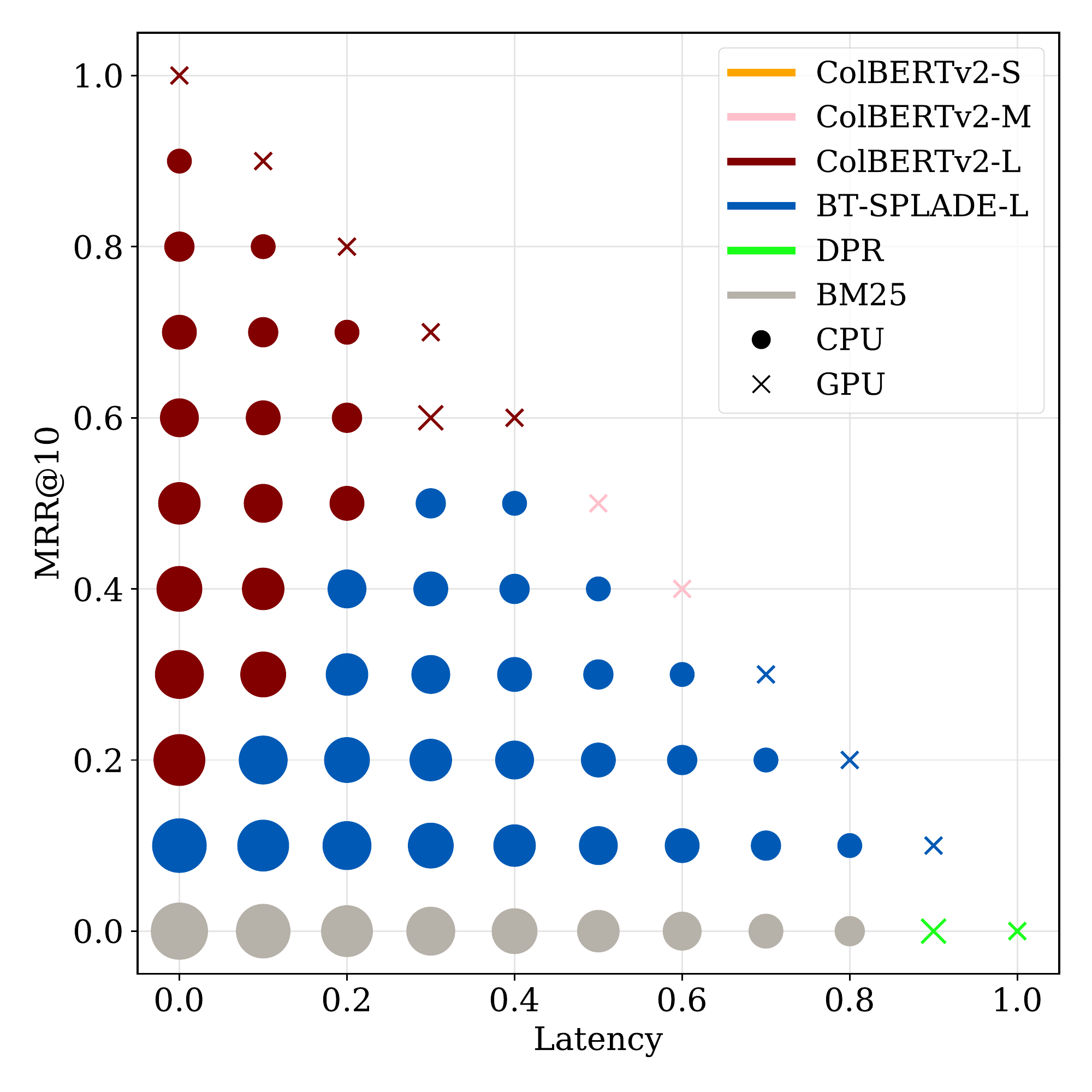}
\caption{XOR-TyDi.}
\label{fig:dynascore-full-xor}
\end{subfigure}
\caption{Exploration of Dynascore weighting schemes. Marker sizes are proportional to Cost weights (large dots represent more-cost-sensitive weightings and thus the most expensive systems are along the diagonal).}
\end{figure*}

For a default weighting, \citeauthor{Ma-etal:2021:Dynaboard} suggest assigning half of the weight to the performance metric and spreading the other half evenly over the other metrics. For our experiments, this leads to
\begin{center}
\{MRR@10: 0.5, Cost: 0.25, Latency: 0.25\}
\end{center}
In Table~\ref{tab:dynascore-leaderboard}, we show what the MS MARCO and XOR-TyDi leaderboards would look like if they were driven by this Dynascore weighting. In both leaderboards, ColBERTv2 variants are the winning systems. This is very decisive for XOR-TyDi. For MS MARCO, ColBERTv2 and SPLADE are much closer overall.

However, this weighting scheme is not the only reasonable choice one could make. Appendix~\ref{app:dynascoring}  presents a range of different leaderboards capturing different relative values. Here, we mention a few highlights. First, if accuracy is very important (e.g., MRR@10: 0.9), then all the ColBERTv2 systems dominate all the others. Second, if we are very cost sensitive, then we could use a weighting \{MRR@10: 0.4, Cost: 0.4, Latency: 0.2\}. In this setting, ColBERTv2-S rises to the top of the leaderboard for MS~MARCO and BT-SPLADE-L is more of a contender. Third, on the other hand, if money is no object, we could use a weighting like \{MRR@10: 0.75, Cost: 0.01, Latency: 0.24\}. This setting justifies using a GPU with COlBERTv2, whereas most other settings do not justify the expense of a GPU for this system. In contrast, a GPU is never justified for BT-SPLADE-L.

To get a holistic picture of how different weightings affect these leaderboards, we conducted a systematic exploration of different weighting vectors. Figure~\ref{fig:dynascore-full} summarizes these findings in terms of the winning system for each setting. The plots depict Latency on the x-axis and Accuracy on the y-axis. The three weights always sum to 1 (Dynascores are normalized), so the Cost value is determined by the other two, as 1.0 -- Accuracy -- Latency. 

The overall picture is clear. For MS~MARCO, a ColBERTv2-M or ColBERTvs-S system is generally the best choice overall assuming Accuracy is the most important value, and ColBERTv2-L is never a winner. In contrast, a BT-SPLADE-L system is generally the best choice where Cost and Latency are much more important than Accuracy. DPR is a winner only where Accuracy is relatively unimportant, and BM25 is a winner only where Accuracy is assigned essentially zero importance. For the out-of-domain XOR-TyDi test, the picture is somewhat different: now ColBERTv2-L is the dominant system, followed by BT-SPLADE-L.

\subsection{Metrics}\label{sec:discussion_metrics}

Here we briefly explore various metrics and their potential role in leaderboard design, beginning with the two that we focused on in our experiments:

\paragraph{Latency} Latency measures the time for a single query to be executed and a result to be returned to the user. Some existing work has measured latency on a single CPU thread to isolate the system performance from potential noise \citep{mackenzie2021wacky,lassance2022efficiency}. While this approach ensures a level playing field for different systems, it fails to reward systems which do benefit from accelerated computation (e.g., on GPUs) or \textit{intra-query} parallelism such as DPR and PLAID ColBERTv2. Therefore, for leaderboards with raw latency as a primary objective, we recommend allowing flexibility in the evaluation hardware to enable the fastest possible submissions. Such flexibility is then subsumed in the dollar cost below.

\paragraph{Dollar cost} Measuring the financial overhead of deploying IR systems is key for production settings. One way to measure cost is to select a particular public cloud instance type and simply multiply the instance rental rate by the time to execute some fixed number of queries, as in Table~\ref{table:post_hoc_leaderboard_estimated_small}. 

\paragraph{Throughput} Throughput measures the total number of queries which can be executed over a fixed time period. Maximizing throughput could entail compromising the average per-query latency in favor of completing a larger volume of queries concurrently. It is important that leaderboards explicitly define the methodology for measuring latency and/or throughput in practice (e.g., in terms of average time to complete one query at a time or average time to complete a batch of 16 queries).

\paragraph{FLOPs} The number of floating point operations (FLOPs) executed by a particular model gives a hardware-agnostic metric for assessing computational complexity. While this metric is meaningful in the context of compute-bound operations such as language modeling~\cite{liu2021towards}, IR systems are often comprised of heterogeneous pipelines where the bottleneck operation may instead be bandwidth-bound~\cite{Santhanam:Khattab-etal:2022}. Therefore we discourage FLOPs as a metric to compete on for IR leaderboards.

\paragraph{Memory usage} IR systems often pre-compute large indexes and load them into memory~\cite{johnson2019billion, khattab2020colbert}, meaning memory usage is an important consideration for determining the minimal hardware necessary to run a given system. In particular, we recommend leaderboard submissions report the index size at minimum as well as the dynamic peak memory usage if possible. The reporting of the dollar cost of each system (i.e., which accounts for the total RAM made available for each system) allows us to quantify the effect of this dimension in practice.
\section{Conclusion} \label{sec:conclusion}

We argued that current benchmarks for information retrieval should adopt multidimensional leaderboards that assess systems based on latency and cost as well as standard accuracy-style metrics. Such leaderboards would likely have the effect of spurring innovation, and lead to more thorough experimentation and more detailed reporting of results in the literature. As a proof of concept, we conducted experiments with four representative IR systems, measuring latency, cost, and accuracy, and showed that this reveals important differences between these systems that are hidden if only accuracy is reported. Finally, we tentatively proposed Dynascoring as a simple, flexible method for creating multidimensional leaderboards in this space.
\section{Limitations} %

We identify two sources of limitations in our work: the range of metrics we consider, and the range of models we explore in our experiments.

Our paper advocates for multidimensional leaderboards. In the interest of concision, we focused on cost and latency as well as system quality. These choices reflect a particular set of values when it comes to developing retrieval models. In \S\ref{sec:discussion_metrics}, we briefly consider a wider range of metrics and highlight some of the values they encode. Even this list is not exhaustive, however. In general, we hope that our work leads to more discussion of the values that should be captured in the leaderboards in this space, and so we do not intend our choices to limit exploration here.

For our post-hoc leaderboard (Table~\ref{table:post_hoc_leaderboard_reported}), we surveyed the literature to find representative systems. We cannot claim that we have exhaustively listed all systems, and any omissions should count as limitations of our work. In particular, we note that we did not consider any re-ranking models, which would consume the top-$k$ results from any of the retrievers we test and produce a re-arranged list. Such models would only add weight to our argument of diverse cost-quality tradeoffs, as re-ranking systems must determine which retriever to re-rank, how many passages to re-rank per query (i.e., setting $k$), and what hardware to use for re-ranking models, which are typically especially accelerator-intensive (i.e., require GPUs or TPUs).

For our experimental comparisons, we chose four models that we take to be representative of broad approaches in this area. However, different choices from within the space of all possibilities might have led to different conclusions. In addition, our experimental protocols may interact with our model choices in important ways. For example, the literature on SPLADE suggests that it may be able to fit its index on machines with 8 or 16 GB of RAM, but our experiments used 32 GB of RAM.

Our hope is merely that our results help encourage the development of leaderboards that offer numerous, fine-grained comparisons from many members of the scientific community, and that these leaderboards come to reflect different values for scoring and ranking such systems as well.

\bibliography{paper,anthology}
\bibliographystyle{acl_natbib}

\newpage
\clearpage
\onecolumn

\appendix

\section*{Supplementary Materials}

\section{Experiment Details}\label{app:experiment_details}

This section provides additional detail for the experiments presented in \S\ref{sec:experimental}.

\paragraph{Datasets} We use the MS MARCO Passage Ranking task unmodified. We use the data from the XOR-Retrieve task (part of XOR-TyDi benchmark), but pre-translate all queries to English. All systems use the same set of pre-translated queries.

\paragraph{Software} We use commit \href{https://github.com/primeqa/primeqa/tree/7b5aa6cc026c5b50bf7bd64e3a7593330d006dac}{7b5aa6c} of PrimeQA and commit \href{https://github.com/naver/splade/tree/d96f5f1710c0684992d05b91e04d7722da474305}{d96f5f1} of SPLADE. We use the provided pip environment files provided by PrimeQA (shared across BM25, DPR, and PLAID ColBERTv2) and SPLADE. The only modification we made to the respective environments was upgrading the PyTorch version in both cases to 1.13. We use Python version 3.9.13 for all experiments.

\paragraph{Hyperparameters} Table~\ref{table:hyperparams} lists the maximum query and passage lengths used for each neural model: \\
\begin{table}[h]
\centering
\begin{tabular}{@{}lrr@{}}
\toprule
\multicolumn{1}{c}{Model} & \multicolumn{1}{c}{|$Q$|} & \multicolumn{1}{c}{|$D$|} \\ \midrule
DPR                       & 32                      & 128                     \\
PLAID ColBERTv2           & 32                      & 300                     \\
BT-SPLADE-Large           & 256                     & 256                     \\ \bottomrule
\end{tabular}
\caption{Maximum query and passage lengths used for each neural model as measured in number of tokens.}
\label{table:hyperparams}
\end{table}

\paragraph{Methodology} We run 10 warm-up iterations for each system to mitigate noise from the initial ramp-up phase. We used Docker containers to ensure precise resource allocations across CPU threads, GPUs, and memory. Our experiments are conducted on AWS instances. The times to instantiate the instance and load model environments are not included in latency calculations.

\paragraph{Model Pre-training and Finetuning} The BM25 model used in our experiments was not pretrained or finetuned for either MSMARCO or XOR-TyDi. Our DPR model used the \textit{facebook/dpr-question\_encoder-multiset-base} and \textit{facebook/dpr-ctx\_encoder-multiset-base} pretrained models and finetunes them on the MSMARCO training set; for XOR-TyDi, our DPR model is not finetuned beyond the original configuration. For BT-SPLADE-Large, we use the \textit{naver/efficient-splade-VI-BT-large-doc} and \textit{naver/efficient-splade-VI-BT-large-query} pretrained models and finetune them on the MSMARCO training set; for XOR-TyDi, we do not finetune them. For PLAID, we use the original model given in \citet{Santhanam:Khattab-etal:2022} and finetune it using the MSMARCO training set; for XOR-TyDi, we do not finetune the model.

\section{Additional Dynascore-Based Leaderboards}\label{app:dynascoring}

\begin{table*}[ht]
    \begin{subtable}{0.48\linewidth}
        \resizebox{\columnwidth}{!}{
            \begin{tabular}{lllr}
\toprule
{} &       System &                     Hardware &  Dynascore \\
\midrule
1  &  ColBERTv2-M &         16 CPU, 32 GB memory &     19.127 \\
2  &  ColBERTv2-S &         16 CPU, 32 GB memory &     19.118 \\
3  &  ColBERTv2-L &         16 CPU, 32 GB memory &     18.857 \\
4  &  ColBERTv2-S &   1 GPU, 1 CPU, 32 GB memory &     18.698 \\
5  &  BT-SPLADE-L &         16 CPU, 32 GB memory &     18.637 \\
6  &  BT-SPLADE-L &          1 CPU, 32 GB memory &     18.616 \\
7  &  ColBERTv2-M &   1 GPU, 1 CPU, 32 GB memory &     18.385 \\
8  &  ColBERTv2-S &          1 CPU, 32 GB memory &     17.912 \\
9  &  BT-SPLADE-L &   1 GPU, 1 CPU, 32 GB memory &     17.839 \\
10 &  ColBERTv2-S &  1 GPU, 16 CPU, 32 GB memory &     17.331 \\
11 &  ColBERTv2-L &   1 GPU, 1 CPU, 32 GB memory &     17.080 \\
12 &  ColBERTv2-M &          1 CPU, 32 GB memory &     17.060 \\
13 &  ColBERTv2-M &  1 GPU, 16 CPU, 32 GB memory &     16.619 \\
14 &  BT-SPLADE-L &  1 GPU, 16 CPU, 32 GB memory &     16.062 \\
15 &  ColBERTv2-L &          1 CPU, 32 GB memory &     15.858 \\
16 &          DPR &         16 CPU, 32 GB memory &     15.635 \\
17 &          DPR &   1 GPU, 1 CPU, 32 GB memory &     15.330 \\
18 &  ColBERTv2-L &  1 GPU, 16 CPU, 32 GB memory &     14.940 \\
19 &          DPR &          1 CPU, 32 GB memory &     14.583 \\
20 &          DPR &  1 GPU, 16 CPU, 32 GB memory &     14.252 \\
21 &         BM25 &           1 CPU, 4 GB memory &      9.263 \\
22 &         BM25 &          1 CPU, 32 GB memory &      9.263 \\
23 &         BM25 &         16 CPU, 32 GB memory &      9.248 \\
24 &         BM25 &          16 CPU, 4 GB memory &      9.246 \\
25 &         BM25 &   1 GPU, 1 CPU, 32 GB memory &      9.072 \\
26 &         BM25 &    1 GPU, 1 CPU, 4 GB memory &      9.049 \\
27 &         BM25 &  1 GPU, 16 CPU, 32 GB memory &      8.565 \\
28 &         BM25 &   1 GPU, 16 CPU, 4 GB memory &      8.551 \\
\bottomrule
\end{tabular}

    }
    \caption{Default weighting per \citealt{Ma-etal:2021:Dynaboard}: \\
    \{Accuracy:~0.5, Cost:~0.25, Latency:~0.25\}.}
    \end{subtable}
    \hfill
    \begin{subtable}{0.48\linewidth}
        \resizebox{\columnwidth}{!}{
            \begin{tabular}{lllr}
\toprule
{} &       System &                     Hardware &  Dynascore \\
\midrule
1  &  ColBERTv2-M &         16 CPU, 32 GB memory &     35.577 \\
2  &  ColBERTv2-L &         16 CPU, 32 GB memory &     35.515 \\
3  &  ColBERTv2-M &   1 GPU, 1 CPU, 32 GB memory &     35.429 \\
4  &  ColBERTv2-S &         16 CPU, 32 GB memory &     35.344 \\
5  &  ColBERTv2-S &   1 GPU, 1 CPU, 32 GB memory &     35.260 \\
6  &  ColBERTv2-M &          1 CPU, 32 GB memory &     35.164 \\
7  &  ColBERTv2-L &   1 GPU, 1 CPU, 32 GB memory &     35.160 \\
8  &  ColBERTv2-S &          1 CPU, 32 GB memory &     35.102 \\
9  &  ColBERTv2-M &  1 GPU, 16 CPU, 32 GB memory &     35.076 \\
10 &  ColBERTv2-S &  1 GPU, 16 CPU, 32 GB memory &     34.986 \\
11 &  ColBERTv2-L &          1 CPU, 32 GB memory &     34.916 \\
12 &  ColBERTv2-L &  1 GPU, 16 CPU, 32 GB memory &     34.732 \\
13 &  BT-SPLADE-L &         16 CPU, 32 GB memory &     34.151 \\
14 &  BT-SPLADE-L &          1 CPU, 32 GB memory &     34.147 \\
15 &  BT-SPLADE-L &   1 GPU, 1 CPU, 32 GB memory &     33.992 \\
16 &  BT-SPLADE-L &  1 GPU, 16 CPU, 32 GB memory &     33.636 \\
17 &          DPR &         16 CPU, 32 GB memory &     28.487 \\
18 &          DPR &   1 GPU, 1 CPU, 32 GB memory &     28.426 \\
19 &          DPR &          1 CPU, 32 GB memory &     28.277 \\
20 &          DPR &  1 GPU, 16 CPU, 32 GB memory &     28.210 \\
21 &         BM25 &           1 CPU, 4 GB memory &     16.813 \\
22 &         BM25 &          1 CPU, 32 GB memory &     16.813 \\
23 &         BM25 &         16 CPU, 32 GB memory &     16.810 \\
24 &         BM25 &          16 CPU, 4 GB memory &     16.809 \\
25 &         BM25 &   1 GPU, 1 CPU, 32 GB memory &     16.774 \\
26 &         BM25 &    1 GPU, 1 CPU, 4 GB memory &     16.770 \\
27 &         BM25 &  1 GPU, 16 CPU, 32 GB memory &     16.673 \\
28 &         BM25 &   1 GPU, 16 CPU, 4 GB memory &     16.670 \\
\bottomrule
\end{tabular}

        }
    \caption{Heavy emphasis on quality: \\
    \{Accuracy:~0.9, Cost:~0.05, Latency:~0.05\}.}
    \end{subtable}

    \vspace{12pt}

    \begin{subtable}{0.48\linewidth}
        \resizebox{\columnwidth}{!}{
            \begin{tabular}{lllr}
\toprule
{} &       System &                     Hardware &  Dynascore \\
\midrule
1  &  ColBERTv2-M &  1 GPU, 16 CPU, 32 GB memory &     29.388 \\
2  &  ColBERTv2-M &   1 GPU, 1 CPU, 32 GB memory &     29.347 \\
3  &  ColBERTv2-M &         16 CPU, 32 GB memory &     29.300 \\
4  &  ColBERTv2-S &  1 GPU, 16 CPU, 32 GB memory &     29.267 \\
5  &  ColBERTv2-S &   1 GPU, 1 CPU, 32 GB memory &     29.259 \\
6  &  ColBERTv2-L &  1 GPU, 16 CPU, 32 GB memory &     29.181 \\
7  &  ColBERTv2-S &         16 CPU, 32 GB memory &     29.172 \\
8  &  ColBERTv2-L &         16 CPU, 32 GB memory &     29.122 \\
9  &  ColBERTv2-L &   1 GPU, 1 CPU, 32 GB memory &     28.961 \\
10 &  BT-SPLADE-L &         16 CPU, 32 GB memory &     28.278 \\
11 &  BT-SPLADE-L &          1 CPU, 32 GB memory &     28.186 \\
12 &  BT-SPLADE-L &   1 GPU, 1 CPU, 32 GB memory &     28.183 \\
13 &  BT-SPLADE-L &  1 GPU, 16 CPU, 32 GB memory &     28.175 \\
14 &  ColBERTv2-S &          1 CPU, 32 GB memory &     28.045 \\
15 &  ColBERTv2-M &          1 CPU, 32 GB memory &     27.422 \\
16 &  ColBERTv2-L &          1 CPU, 32 GB memory &     26.407 \\
17 &          DPR &         16 CPU, 32 GB memory &     23.634 \\
18 &          DPR &   1 GPU, 1 CPU, 32 GB memory &     23.622 \\
19 &          DPR &  1 GPU, 16 CPU, 32 GB memory &     23.586 \\
20 &          DPR &          1 CPU, 32 GB memory &     22.708 \\
21 &         BM25 &         16 CPU, 32 GB memory &     13.958 \\
22 &         BM25 &          16 CPU, 4 GB memory &     13.958 \\
23 &         BM25 &          1 CPU, 32 GB memory &     13.952 \\
24 &         BM25 &           1 CPU, 4 GB memory &     13.945 \\
25 &         BM25 &   1 GPU, 1 CPU, 32 GB memory &     13.944 \\
26 &         BM25 &    1 GPU, 1 CPU, 4 GB memory &     13.936 \\
27 &         BM25 &  1 GPU, 16 CPU, 32 GB memory &     13.931 \\
28 &         BM25 &   1 GPU, 16 CPU, 4 GB memory &     13.930 \\
\bottomrule
\end{tabular}

        }
        \caption{Cost is not a concern, and low latency is key:\\ 
        \{Accuracy:~0.75, Cost:~0.01, Latency:~0.24\}.}
    \end{subtable}
    \hfill
    \begin{subtable}{0.48\linewidth}
        \resizebox{\columnwidth}{!}{
            \begin{tabular}{lllr}
\toprule
{} &       System &                     Hardware &  Dynascore \\
\midrule
1  &  ColBERTv2-S &         16 CPU, 32 GB memory &     15.138 \\
2  &  ColBERTv2-M &         16 CPU, 32 GB memory &     15.108 \\
3  &  BT-SPLADE-L &          1 CPU, 32 GB memory &     14.851 \\
4  &  ColBERTv2-L &         16 CPU, 32 GB memory &     14.820 \\
5  &  BT-SPLADE-L &         16 CPU, 32 GB memory &     14.806 \\
6  &  ColBERTv2-S &   1 GPU, 1 CPU, 32 GB memory &     14.375 \\
7  &  ColBERTv2-S &          1 CPU, 32 GB memory &     14.146 \\
8  &  ColBERTv2-M &   1 GPU, 1 CPU, 32 GB memory &     13.855 \\
9  &  BT-SPLADE-L &   1 GPU, 1 CPU, 32 GB memory &     13.584 \\
10 &  ColBERTv2-M &          1 CPU, 32 GB memory &     13.363 \\
11 &          DPR &         16 CPU, 32 GB memory &     12.451 \\
12 &  ColBERTv2-L &          1 CPU, 32 GB memory &     12.278 \\
13 &  ColBERTv2-S &  1 GPU, 16 CPU, 32 GB memory &     12.132 \\
14 &  ColBERTv2-L &   1 GPU, 1 CPU, 32 GB memory &     12.055 \\
15 &          DPR &   1 GPU, 1 CPU, 32 GB memory &     11.962 \\
16 &          DPR &          1 CPU, 32 GB memory &     11.537 \\
17 &  ColBERTv2-M &  1 GPU, 16 CPU, 32 GB memory &     10.932 \\
18 &  BT-SPLADE-L &  1 GPU, 16 CPU, 32 GB memory &     10.687 \\
19 &          DPR &  1 GPU, 16 CPU, 32 GB memory &     10.231 \\
20 &  ColBERTv2-L &  1 GPU, 16 CPU, 32 GB memory &      8.365 \\
21 &         BM25 &           1 CPU, 4 GB memory &      7.408 \\
22 &         BM25 &          1 CPU, 32 GB memory &      7.401 \\
23 &         BM25 &         16 CPU, 32 GB memory &      7.371 \\
24 &         BM25 &          16 CPU, 4 GB memory &      7.369 \\
25 &         BM25 &   1 GPU, 1 CPU, 32 GB memory &      7.095 \\
26 &         BM25 &    1 GPU, 1 CPU, 4 GB memory &      7.065 \\
27 &         BM25 &  1 GPU, 16 CPU, 32 GB memory &      6.278 \\
28 &         BM25 &   1 GPU, 16 CPU, 4 GB memory &      6.256 \\
\bottomrule
\end{tabular}

        }
    \caption{Cost is a very significant concern:\\ \{Accuracy:~0.4, Cost:~0.4, Latency:~0.2\}.}
    \end{subtable}   
\caption{Dynascores for MS MARCO, for different weightings of the metrics.}
\label{tab:dynascores-msmarco}
\end{table*}

\newpage

\begin{table*}[ht]
    \begin{subtable}{0.48\linewidth}
        \resizebox{\columnwidth}{!}{
            \begin{tabular}{lllr}
\toprule
{} &       System &                     Hardware &  Dynascore \\
\midrule
1  &  ColBERTv2-L &         16 CPU, 64 GB memory &     21.241 \\
2  &  BT-SPLADE-L &         16 CPU, 64 GB memory &     21.119 \\
3  &  ColBERTv2-M &         16 CPU, 64 GB memory &     21.063 \\
4  &  BT-SPLADE-L &          1 CPU, 64 GB memory &     20.753 \\
5  &  ColBERTv2-M &  1 GPU, 16 CPU, 64 GB memory &     20.255 \\
6  &  BT-SPLADE-L &  1 GPU, 16 CPU, 64 GB memory &     20.123 \\
7  &  ColBERTv2-M &   1 GPU, 1 CPU, 64 GB memory &     19.904 \\
8  &  ColBERTv2-L &  1 GPU, 16 CPU, 64 GB memory &     19.700 \\
9  &  ColBERTv2-S &         16 CPU, 64 GB memory &     19.649 \\
10 &  BT-SPLADE-L &   1 GPU, 1 CPU, 64 GB memory &     19.380 \\
11 &  ColBERTv2-S &  1 GPU, 16 CPU, 64 GB memory &     19.157 \\
12 &  ColBERTv2-L &   1 GPU, 1 CPU, 64 GB memory &     19.123 \\
13 &  ColBERTv2-S &   1 GPU, 1 CPU, 64 GB memory &     18.723 \\
14 &  ColBERTv2-S &          1 CPU, 64 GB memory &     15.934 \\
15 &         BM25 &          1 CPU, 64 GB memory &     12.635 \\
16 &         BM25 &         16 CPU, 64 GB memory &     12.630 \\
17 &         BM25 &  1 GPU, 16 CPU, 64 GB memory &     11.847 \\
18 &         BM25 &   1 GPU, 1 CPU, 64 GB memory &     11.794 \\
19 &  ColBERTv2-M &          1 CPU, 64 GB memory &     11.563 \\
20 &  ColBERTv2-L &          1 CPU, 64 GB memory &      7.708 \\
21 &          DPR &   1 GPU, 1 CPU, 64 GB memory &      7.452 \\
22 &          DPR &  1 GPU, 16 CPU, 64 GB memory &      7.386 \\
23 &          DPR &         16 CPU, 64 GB memory &      7.188 \\
24 &          DPR &          1 CPU, 64 GB memory &      5.442 \\
\bottomrule
\end{tabular}

    }
    \caption{Default weighting per \citealt{Ma-etal:2021:Dynaboard}: \\
    \{Accuracy:~0.5, Cost:~0.25, Latency:~0.25\}.}
    \end{subtable}
    \hfill
    \begin{subtable}{0.48\linewidth}
        \resizebox{\columnwidth}{!}{
            \begin{tabular}{lllr}
\toprule
{} &       System &                     Hardware &  Dynascore \\
\midrule
1  &  ColBERTv2-L &         16 CPU, 64 GB memory &     42.200 \\
2  &  ColBERTv2-L &  1 GPU, 16 CPU, 64 GB memory &     41.892 \\
3  &  ColBERTv2-L &   1 GPU, 1 CPU, 64 GB memory &     41.777 \\
4  &  ColBERTv2-M &         16 CPU, 64 GB memory &     40.557 \\
5  &  ColBERTv2-M &  1 GPU, 16 CPU, 64 GB memory &     40.395 \\
6  &  ColBERTv2-M &   1 GPU, 1 CPU, 64 GB memory &     40.325 \\
7  &  ColBERTv2-L &          1 CPU, 64 GB memory &     39.494 \\
8  &  BT-SPLADE-L &         16 CPU, 64 GB memory &     39.048 \\
9  &  BT-SPLADE-L &          1 CPU, 64 GB memory &     38.975 \\
10 &  BT-SPLADE-L &  1 GPU, 16 CPU, 64 GB memory &     38.849 \\
11 &  BT-SPLADE-L &   1 GPU, 1 CPU, 64 GB memory &     38.700 \\
12 &  ColBERTv2-M &          1 CPU, 64 GB memory &     38.657 \\
13 &  ColBERTv2-S &         16 CPU, 64 GB memory &     37.362 \\
14 &  ColBERTv2-S &  1 GPU, 16 CPU, 64 GB memory &     37.263 \\
15 &  ColBERTv2-S &   1 GPU, 1 CPU, 64 GB memory &     37.177 \\
16 &  ColBERTv2-S &          1 CPU, 64 GB memory &     36.619 \\
17 &         BM25 &          1 CPU, 64 GB memory &     23.599 \\
18 &         BM25 &         16 CPU, 64 GB memory &     23.598 \\
19 &         BM25 &  1 GPU, 16 CPU, 64 GB memory &     23.441 \\
20 &         BM25 &   1 GPU, 1 CPU, 64 GB memory &     23.431 \\
21 &          DPR &   1 GPU, 1 CPU, 64 GB memory &     15.010 \\
22 &          DPR &  1 GPU, 16 CPU, 64 GB memory &     14.997 \\
23 &          DPR &         16 CPU, 64 GB memory &     14.958 \\
24 &          DPR &          1 CPU, 64 GB memory &     14.608 \\
\bottomrule
\end{tabular}

        }
    \caption{Heavy emphasis on quality: \\
    \{Accuracy:~0.9, Cost:~0.05, Latency:~0.05\}.}
    \end{subtable}

    \vspace{12pt}

    \begin{subtable}{0.48\linewidth}
        \resizebox{\columnwidth}{!}{
            \begin{tabular}{lllr}
\toprule
{} &       System &                     Hardware &  Dynascore \\
\midrule
1  &  ColBERTv2-L &  1 GPU, 16 CPU, 64 GB memory &     34.073 \\
2  &  ColBERTv2-L &   1 GPU, 1 CPU, 64 GB memory &     33.859 \\
3  &  ColBERTv2-L &         16 CPU, 64 GB memory &     33.385 \\
4  &  ColBERTv2-M &  1 GPU, 16 CPU, 64 GB memory &     33.149 \\
5  &  ColBERTv2-M &   1 GPU, 1 CPU, 64 GB memory &     33.020 \\
6  &  ColBERTv2-M &         16 CPU, 64 GB memory &     32.609 \\
7  &  BT-SPLADE-L &         16 CPU, 64 GB memory &     32.076 \\
8  &  BT-SPLADE-L &  1 GPU, 16 CPU, 64 GB memory &     32.036 \\
9  &  BT-SPLADE-L &   1 GPU, 1 CPU, 64 GB memory &     31.752 \\
10 &  BT-SPLADE-L &          1 CPU, 64 GB memory &     31.718 \\
11 &  ColBERTv2-S &  1 GPU, 16 CPU, 64 GB memory &     30.689 \\
12 &  ColBERTv2-S &   1 GPU, 1 CPU, 64 GB memory &     30.532 \\
13 &  ColBERTv2-S &         16 CPU, 64 GB memory &     30.239 \\
14 &  ColBERTv2-S &          1 CPU, 64 GB memory &     26.788 \\
15 &  ColBERTv2-M &          1 CPU, 64 GB memory &     23.835 \\
16 &  ColBERTv2-L &          1 CPU, 64 GB memory &     20.881 \\
17 &         BM25 &         16 CPU, 64 GB memory &     19.276 \\
18 &         BM25 &          1 CPU, 64 GB memory &     19.264 \\
19 &         BM25 &  1 GPU, 16 CPU, 64 GB memory &     19.258 \\
20 &         BM25 &   1 GPU, 1 CPU, 64 GB memory &     19.243 \\
21 &          DPR &   1 GPU, 1 CPU, 64 GB memory &     12.305 \\
22 &          DPR &  1 GPU, 16 CPU, 64 GB memory &     12.277 \\
23 &          DPR &         16 CPU, 64 GB memory &     11.558 \\
24 &          DPR &          1 CPU, 64 GB memory &      9.913 \\
\bottomrule
\end{tabular}

        }
        \caption{Cost is not a concern, and low latency is key:\\ 
        \{Accuracy:~0.75, Cost:~0.01, Latency:~0.24\}.}
    \end{subtable}
    \hfill
    \begin{subtable}{0.48\linewidth}
        \resizebox{\columnwidth}{!}{
            \begin{tabular}{lllr}
\toprule
{} &       System &                     Hardware &  Dynascore \\
\midrule
1  &  BT-SPLADE-L &         16 CPU, 64 GB memory &     16.853 \\
2  &  ColBERTv2-L &         16 CPU, 64 GB memory &     16.832 \\
3  &  ColBERTv2-M &         16 CPU, 64 GB memory &     16.743 \\
4  &  BT-SPLADE-L &          1 CPU, 64 GB memory &     16.565 \\
5  &  ColBERTv2-S &         16 CPU, 64 GB memory &     15.638 \\
6  &  BT-SPLADE-L &  1 GPU, 16 CPU, 64 GB memory &     15.259 \\
7  &  ColBERTv2-M &  1 GPU, 16 CPU, 64 GB memory &     14.954 \\
8  &  ColBERTv2-M &   1 GPU, 1 CPU, 64 GB memory &     14.491 \\
9  &  ColBERTv2-S &  1 GPU, 16 CPU, 64 GB memory &     14.444 \\
10 &  BT-SPLADE-L &   1 GPU, 1 CPU, 64 GB memory &     14.291 \\
11 &  ColBERTv2-S &   1 GPU, 1 CPU, 64 GB memory &     13.871 \\
12 &  ColBERTv2-L &  1 GPU, 16 CPU, 64 GB memory &     13.714 \\
13 &  ColBERTv2-L &   1 GPU, 1 CPU, 64 GB memory &     12.958 \\
14 &  ColBERTv2-S &          1 CPU, 64 GB memory &     12.566 \\
15 &         BM25 &          1 CPU, 64 GB memory &     10.088 \\
16 &         BM25 &         16 CPU, 64 GB memory &     10.069 \\
17 &  ColBERTv2-M &          1 CPU, 64 GB memory &      8.843 \\
18 &         BM25 &  1 GPU, 16 CPU, 64 GB memory &      8.806 \\
19 &         BM25 &   1 GPU, 1 CPU, 64 GB memory &      8.731 \\
20 &          DPR &         16 CPU, 64 GB memory &      5.669 \\
21 &  ColBERTv2-L &          1 CPU, 64 GB memory &      5.582 \\
22 &          DPR &   1 GPU, 1 CPU, 64 GB memory &      5.450 \\
23 &          DPR &  1 GPU, 16 CPU, 64 GB memory &      5.368 \\
24 &          DPR &          1 CPU, 64 GB memory &      4.244 \\
\bottomrule
\end{tabular}

        }
    \caption{Cost is a very significant concern:\\ \{Accuracy:~0.4, Cost:~0.4, Latency:~0.2\}.}
    \end{subtable}
\caption{Dynascores for XOR-TyDi, for different weightings of the metrics.}
\label{tab:dynascores-xor}
\end{table*}

\end{document}